\documentstyle[floats,aps,eqsecnum]{revtex}
\begin{document}
\draft
\catcode`\@=11
\catcode`\@=12
\twocolumn[\hsize\textwidth\columnwidth\hsize\csname%
@twocolumnfalse\endcsname
\title{The Microscopic Picture of  Chiral Luttinger Liquid:
Composite Fermion Theory of Edge States}
\author{Yue Yu, Wenjun Zheng and Zhongyuan Zhu}
\address{Institute of Theoretical Physics, Academia Sinica, Beijing 100080,
China}

\maketitle
\begin{abstract}
 
We derive a microscopic theory of the composite fermions describing the 
low-lying edge excitations in the fractional quantum 
Hall liquid. Using the composite fermion
transformation, one finds that the edge states of the $\nu=1/m$ 
system in a disc sample are described by, in one dimensional limit, the 
Calogero-Sutherland model with other interactions between the composite 
fermions as perturbations. It is shown that a large class of short-range 
interactions renormalize only the Fermi velocity while the
exponent $g=\nu=1/m$ is invariant under the condition of chirality.
By taking the sharp edge potential into
account, we obtain a microscopic
justification of the chiral Luttinger liquid model of 
the fractional quantum Hall edge states. The approach applied to the 
$\nu=1/m$ system can be generalized to the other edge states with odd 
denominator filling factors.     

\end{abstract}

]


\section{INTRODUCTION}

The study of edge excitations of fractional quantum Hall effect (FQHE) 
has evoked considerable interests in the past decade. Compared with
the edge states of integer quantum Hall effect (IQHE), which 
are well understood to be a chiral Fermi liquid \cite{Halp}, the description of
FQHE is a strongly 
correlated many-body problem. It is now believed that the bulk property of a 
FQH liquid can be described by an incompressible quantum liquid with only 
gapful excitations \cite{Lau}, while the gapless ones exist along the edge. 
In order to understand the edge excitations many attempts have been made 
\cite{Rez,ECS,CFP,Wen,DO}. From a macroscopic point of view, Wen suggests that 
the FQHE edge with $\nu=1/m$ ($m$ is an odd integer) is actually a 
chiral Luttinger liquid (CLL) with its characteristic exponent $g=\nu$
as a topological index which is invariant to  perturbations 
like the interactions between electrons\cite{Wen}. In deriving the 
macroscopic theory, one starts from a few general assumptions, such as the 
gauge invariance of the system and arrives at the 
macroscopic effective field theory without being concerned with the details of 
the microscopic interactions. However, a microscopic justification of the 
effective theory is still lacking, despite some recent efforts which make 
use of collective field method to give a random phase approximation solution 
to the equations of motion \cite{DO}. Recently, a mean-field 
consideration of the fermion Chern-Simons theory of the edge states has been 
presented \cite{CFP}. Meanwhile, several 
authors \cite{ECS} point out the possible relations
between the edge states of FQHE and the Calogero-Sutherland model (CSM) 
\cite{CS}.

Recently, two of us \cite{Yu} give a derivation of 
a microscopic model of the composite fermion (CF) \cite{Jain} 
type excitations at the FQHE 
edge with an odd denominator filling factor $\nu=1/m$ and relates the model 
to the CSM in the one-dimensional (1-d) limit. In this paper, we present the 
details and some unpublished results ,and give a 
generalization to the other situations with odd denominator filling factors. 
We start from the general Hamiltonian for a two-dimensional system of 
interacting electrons in a strong perpendicular magnetic field and bound into 
some specific geometry like a disc. After a CF type of anyon \cite{LMW} 
transformation \cite{KHZ,LF,KZ,HLR,RS,WY}, we obtain a Hamiltonian whose 
ground state wave function can be simply written out. By extracting 
the excitation part from the total wave function, we arrive at a form 
of Hamiltonian which is shown to be equivalent to the corresponding Hamiltonian
of the CSM in the 1-d limit. 
We then take into account the influence of the interactions between CFs. 
By using the analog of the 1-d two-body Schrodinger equation to the 
wave equation satisfied by the radial wave function of a particle in a
three-dimensional centrally 
symmetric field, we demonstrate that under the condition of chirality the 
topological exponent $g=1/m$ is 
indeed robust against the perturbation of a class of short-range interactions
like the $\delta$-function interaction and the pseudopotentials \cite{psudo}
while the Fermi velocity can be altered. It is emphasized that the low-lying 
excitations of
the CSM are described by the $c=1$ conformal field theory (CFT) with its 
compactified radius ${\cal R}=1/\sqrt{m}$ \cite{KY,Wu}, which gives a profound 
reason for its low-energy fixed point of a Luttinger liquid.  
The bare Coulomb interaction will affect the exponent because of its 
infrared divergence. However, it is expected that the 1-d Coulomb interaction
is actually drastically renormalized or partially screened by either the bulk 
electrons or the experimental devices such as a metal electrode.
We consider the case of  drastically renormalized Coulomb interaction
and show that the exponent $g$ is not changed while a branch of the 1-d plasmon
excitations appears, which agrees with the CLL's results of the influence
of the Coulomb interaction \cite{Wen1}.

The chirality of the edge excitations is a basic assumption of the macroscopic
effective theory like the CLL. However, the microscopic understanding of it is 
still
elusive. Recently,  
Chklovskii and Halperin try to show it in the CF picture
beyond the mean-field theory \cite{Ch}. In this paper, 
we derive the chirality from the microscopic point of view in the
framework of the present theory. Since the edge CFs are confined in a very
narrow strip near the boundary of the sample, the single particle picture
will work for the radial wave functions except that the wave vectors of the CFs
are strongly coupled through the asymptotic Bethe ansatz equations of the CSM. 
Furthermore, one finds that the left-moving sector of the excitations of the 
CSM is suppressed because the magnetic field breaks the left and 
right moving symmetry of the CSM.  
The sound wave velocity of the theory is renormalized after turning on an 
external electric field and the increment of it agrees with the drift velocity 
of CF in the
effective magnetic field. However, Haldane's velocity relations of the 
Luttinger liquid  \cite{Hald1} are invariant, which implies that the 
low-lying excitations of the system are still governed by the $c=1$ CFT 
without changing the compactified radius . 
In this sense, we finally arrive at the justification of the CLL
theory of the edge excitations for the $\nu=1/m$ FQHE from the microscopic 
point of view.

The approach applied to the edge states with $\nu=1/m$ can be 
generalized to the edge states with other odd denominator filling factors ,where
a $K$-matrix, which is employed by Wen and Zee in the effective field theory of 
FQHE \cite{WZ}, is used to characterise the exponents of the multi-branch CLL.

This paper is organized as follows: In Sec. II,  we give a detailed 
derivation of the microscopic model of the FQHE edge states with $\nu=1/m$
and relate the model to the CSM  whose low-lying excitations are governed by
the $c=1$ CFT. In Sec. III, we discuss the influences of the interactions 
between edge CFs and show that the topological exponent is not renormalized by
some short-range interactions because of chirality while the Coulomb interaction 
leaves some more subtle
effects to be clarified. In Sec. IV, the radial degree of freedom is included
and we provide the microscopic justification of  the CLL theory 
of FQHE edge modes. A generalization to other filling factors 
is presented in Sec. V. We conclude the paper in the final section.

\section {DERIVATION OF THE MICROSCOPIC EDGE MODEL}

We start from the following Hamiltonian which describes a system of 
two dimensional interacting electrons with their spins polarized by
a strong magnetic field,
\begin{eqnarray}
H_{\rm e}&=&\sum_{\alpha=1}^N\frac{1}{2m_b}[\vec{p}_\alpha
+\frac{e}{c}\vec{A}(\vec{r}_\alpha)]^2
+\sum_{\alpha<\beta}V(\vec{r}_\alpha-\vec{r}_\beta)\\ \nonumber
&+&\sum_\alpha U(\vec{r}_\alpha),
\label{HO}
\end{eqnarray}
where $V(\vec {r})$ represents the interactions between electrons,
which will be specified later. $U(\vec{r})$ is an external potential that 
serves as a confinement to hold the electrons in some area of certain 
geometry, say a disc of radius R as it is assumed throughout this paper.
Furthermore, we assume the potential is sharp enough at the boundary of the 
sample to avoid complex structure of the edge states while  
it is smooth away from the edge such that it has no influences on the bulk 
of Hall liquid. $m_b$ is the band mass of the electron and $\vec A$ is 
the vector potential providing an external magnetic field normal to the plane.

The composite particle transformation will bring us to a good
starting point to deal with the FQHE as many successful investigations have
told us \cite{KHZ,LF,KZ,HLR}.    
In order to change into a CF picture, we can perform an anyon transformation 
which reads
\begin{equation}
\Psi_e(z_1,...,z_N) = \prod_{\alpha<\beta}\biggl[\frac{z_\alpha-z_\beta}
{|z_\alpha-z_\beta|}\biggr]^{\tilde\phi} 
\Psi_c(z_1,...,z_N),
\label{CBT}
\end{equation}
where  $\Psi_e$  gives the electronic wave function, while $\Psi_c$ is the 
wave function of CFs. $\tilde\phi$ is the number of flux quanta attached to 
one electron, which is 
an even number for CF. After this transformation, we 
obtain the following Hamiltonian,
\begin{eqnarray}
H_{\rm cf}&=&\sum_{i=1}^{N}[\vec{p}_i+\vec{A}(\vec{r}_i)
+\vec{a}(\vec{r}_i)]^2\\ \nonumber
&+&\sum_{i<j}V(\vec{r}_i-\vec{r}_j)+
\sum_i U_{\rm }(\vec{r}_i),
\end{eqnarray}
where
\begin{equation}
\vec{a}(\vec{r}_{i})=\tilde\phi\sum_{j\not=i}
\frac{\hat{z}\times (\vec{r}_{i}-\vec{r}_{j})}
{|\vec{r}_{i}-\vec{r}_{j}|^2},
\end{equation}
is a statistical gauge potential responsible for the bound flux. In this
CF Hamiltonian, we have replaced the electron band mass $m_b$
by a phenomenological effective mass $m^*$ of  the CF and use the unit
$\hbar=2m^*=e/c=1$ in what follows.

This Hamiltonian can be rewritten in 
a more suggestive form which reads
\begin{eqnarray}
H_c&=&\sum_{i=1}^{N}\biggl[(\frac{-i}{r_i}\frac{\partial}{\partial \phi_i}+
      \frac{\partial E_i'}{\partial r_i})^2+ 
       (-i\frac{\partial}{\partial r_i}-
      \frac{1}{r_i}\frac{\partial E_i'}{\partial \phi_i})^2\\ \nonumber
      &-&\frac{1}{r_i}\frac{\partial}{\partial r_i}
       +\frac{i}{r_i^2}\frac{\partial E_i'}{\partial 
\phi_i}\biggr]+\sum_{i<j}V(\vec{r}_i-\vec{r}_j)+
\sum_i U_{\rm }(\vec{r}_i),    
\end{eqnarray}
where  
\begin{eqnarray}
    E_i&=&\sum_{j}{'}E_{ij} \\ \nonumber
       &=&\tilde\phi \sum_{j}{'} \ln|\vec{r_i}-\vec{r_j}| \\ \nonumber
    E_i'&=&\sum_{j}{'}E_{ij}' \\ \nonumber
              &=&\sum_{j}{'}(E_{ij}-\frac{\beta r_i^2}{N-1} )\\ \nonumber
              &=&E_i-\beta r_i^2,    
\end{eqnarray}
and $\beta=\frac{|B|}{4}$.      
Here we suppose $ B<0$. 
This form reminds us of a two dimensional plasma with $E_{ij}$ closely 
resembling the 2d Coulomb interaction potential.
The analog has been very useful to numeric calculations of FQHE.  
     
Next we conduct a non-unitary transformation of the Hamiltonian $H_c$ as 
follows \cite{RS,WY},
   
\begin{eqnarray}
\Psi_c&=&\exp{(\sum_{i<j}E_{ij}')} \Phi \\
      &=&\exp{(-\sum_{i}\beta r_i^2)} \prod_{i<j}^{} |z_i-z_j|^{\tilde\phi 
}\Phi,
\end{eqnarray}
\begin{eqnarray}
H'&=&\exp{(\sum_{i<j}^{}-E_{ij}{'})}H_c 
\exp{(\sum_{i<j}^{}E_{ij}{'})}\\ \nonumber
        &=&-4\sum_{i=1}^{N}\frac{\partial}{\partial z_i} 
            \frac{\partial}{\partial \bar{z_i}} - 
\sum_{i=1}^{N}\{ 4\tilde\phi\sum_{j} {'} 
\frac{1}{z_i-z_j}-2|B|\bar{z_i}\}\frac{\partial}{\partial \bar{z_i}}\\ 
\nonumber
         &+&\sum_{i<j}V(\vec{r}_i-\vec{r}_j)+\sum_i U_{\rm }(\vec{r}_i).
\end{eqnarray}     
            
Although the Hamiltonian $H'$ so obtained is not Hermitian in itself, 
it does provide us with some insights of the ground state wave function. 
Obviously, any non-singular complex function of $z_i$ gives an eigenstate of 
$H'$, which actually represents its ground state at least in case of the 
CF picture in the sense of mean field, that is
\begin{eqnarray}
\Phi_g&=&f(z_1,..z_N).
\end{eqnarray}  
If we do not consider the interaction, the ground state is highly degenerate.
However, the interaction breaks the degeneracy and the groundstate has to 
maintain 
a minimal angular momentum as Laughlin has suggested 
in the presentation of his famous 
wave functions. For the CF system, the ground state reads
\begin{equation}
f(z_1,...z_N)=\prod_{i<j}^{}(z_i-z_j).
\end{equation}
Thus, we arrive at an effective Hamiltonian by expressing the wave function 
$\Phi$ as a product of the ground state wave function
$ \Phi_g $ and an extra part $ \Phi'' $, and consider only the 
Hamiltonian corresponding to the latter, which is 
\begin{eqnarray}
H''&=& \Phi_g^{-1}H'\Phi_g\\ \nonumber
                &=&-4\sum_{i=1}^{i=N}\frac{\partial}{\partial z_i} 
            \frac{\partial}{\partial \bar{z_i}} \\ \nonumber
&-& \sum_{i=1}^{i=N}\{ 4m
            \sum_{j} {'} 
\frac{1}{z_i-z_j}-2|B|\bar{z_i}\}\frac{\partial}{\partial \bar{z_i}} \\   
\nonumber
            &+&\sum_{i<j}V(\vec{r}_i-\vec{r}_j)+\sum_i U_{\rm }(\vec{r}_i)    
\end{eqnarray}

If we reverse the above transformations from $H''$ to $H_e$,
we will arrive at the ground state wave function 
\begin{equation}
\Psi_{eg}(z_1,...z_N)=\exp{(-\sum_{i}^{}\beta r_i^2)} \prod_{i<j}^{} 
(z_i-z_j)^{m},
\end{equation} 
which is precisely the well-known wave function first proposed by 
Laughlin.

Based on the above understanding of the CF groundstate wave function, we now
turn to the edge states theory of the CF excitations.
The partition function of the system is given by
\begin{eqnarray}
Z&=&\sum_{N^e} C^{N^e}_{N}\int_{\partial} d^2z_1....d^2z_{N^e}
\int_{B}d^2z_{N^e+1}...d^2z_N\\ \nonumber
&\times&\biggl(\sum_\delta |\Psi_\delta|^2 e^{-\beta 
(E_\delta+E_g)}+\sum_\gamma|\Psi_\gamma
|^2e^{-\beta (E_\gamma+E_g)}\biggr),
\label{pf}
\end{eqnarray} 
where we have divided the sample into the edge $\partial$ and the bulk $B$. 
$E_g$ is the ground state energy and $E_\delta$ are the low-lying gapless 
excitation energies with $\delta$ being the excitation  branch index.
$E_\gamma$ are the gapful excitation energies. At $\nu=1/\tilde\phi$, the 
low-lying excitations are everywhere in the sample and we do not consider this
case here. We are interested in the case $\nu=\frac{1}{\tilde\phi+1}=1/m$,
where the bulk states are gapful. The low-lying excitations are confined in 
the edge of the sample. For convenience, we consider a disc geometry sample
here. The edge potential is postulated with a sharp shape. The advantage
of the CF picture is we have a manifestation that the FQHE
of the electrons in the external field $B$ could be understood as the IQHE of 
the CFs in the effective field $B^*$ defined by $B^*\nu^*=B\nu$.
For the present case, $B^*=B/m$ and $\nu^*=1$. The energy gap in the bulk
is of the order $\hbar \omega_c^*$ with the effective cyclotron frequency
$\omega_c^*=\frac{eB^*}{m^* c}$ ($m^*$ is the effective mass of the CF). 
Hereafter, we use the unit $\hbar=e/c=2m^*=1$ except the explicit expressions.
By the construction 
of the CF, the FQHE of the electrons can be described by  
the IQHE of the CFs \cite{Jain} while the electrons in the $\nu=
1/\tilde\phi$ field could be thought as the CFs in a zero effective field.
Thus, a Fermi-liquid like theory could be used \cite{HLR} and we 
have a set of CF-type quasiparticles. Applying the single
particle picture, which Halperin used to analyze the edge excitations of
the IQHE of the electrons, to the edge excitations of
the CFs, one could have a microscopic theory of the quasiparticles at the 
edge. In the low-temperature limit, 
the domination states contributing to the partition function 
are those states that the lowest Landau level of the CF-type excitations  
is fully filled in the bulk but only allow the edge CF-type excitations
to be gapless because the gap is shrinked in the edge due to the 
sharp edge potential. The other states with their energy $E_\gamma+E_g$ open 
a gap at least in the order of $\hbar \omega^*_c$ to the ground state. 
In the low-temperature limit, $k_BT\ll \hbar \omega^*_c$, the effective 
partition function is
\begin{eqnarray}
Z&\simeq&\sum_{\delta, N^e}C_N^{N^e} \int_{\partial} d^2z_1...d^2z_{N^e}
|\Psi_{e,\delta}|^2 e^{-\beta (E_\delta(N^e)+E_{g,b})}\\ \nonumber
&=&\sum_{N^e}C_N^{N^e}{\rm Tr_{(edge)}}e^{-\beta (H_e+E_{g,b})},
\label{apf}
\end{eqnarray}
where the trace runs over the low-lying set of the quantum state space for 
a fixed $N_e$ and, according to the single particle picture, 
$\Psi_{e,\delta}$ are the edge many-quasiparticle wave functions
. $E_\delta(N^e)$ is the eigen energy of the edge quasiparticle excitations
and $E_{g,b}$ is the
bulk state contribution to the ground state energy.

For a disc sample, the edge CFs are restricted in a circular 
strip near the boundary with its width  $\delta r_i<<R $ where $R$ is the radius 
of the disc. Suppose we have $N_e$ CFs at the edge,
while the remaining $N-N_e$ are localized inside the bulk. 
Since we are only 
interested in low energy excitations of the quantum Hall liquid, and the bulk 
CFs contribute no gapless elementary excitations, we can focus our attention 
on the edge CFs by separating them from the bulk in Hamiltonian $H_c$,
and treating the interactions with bulk ones as an `external potential' which 
includes both vector potential and scalar one. The former supplies, in the
mean-field approximation, an additional magnetic field that on average reduces 
the external magnetic field from B to $B^*$, that is
\begin{eqnarray*}
       E_i^e&=&\biggl \langle\sum_{j\in 
bulk} {'}E_{ij}\biggr\rangle_{bulk}+\sum_{j\in edge} {'}E_{ij}
  -\beta r_i^2\\
                 &=&\sum_{j\in edge}{'}E_{ij}-\beta^* r_i^2
\end{eqnarray*}
where $ \beta^*=\frac{|B^*|}{4}$ and $B^*=B/m$. 
     
The scalar potential can partly screen the Coulomb interaction between 
the edge particles. If the magnetic field is absent, the screening is caused
by  `mirror positive charges' that provide the scalar potential to slowly moving 
edge CFs so as to reduce the 1/r interaction to one with shorter range like 
dipole-dipole interaction. However, the effective magnetic field forces the 
CF's into cyclotron motion and the screening may be weakened. Even so, we still
expect the bare Coulomb interaction between the edge CFs to be dramatically 
renormalized or screened
to a shorter range interaction by considering the effect of the experimental 
devices such as a metal electrode. 
We represent the modified interaction potential 
as $V_{\rm sc}(\vec{r}_i-\vec{r}_j)$ and the modified form of 
the edge potential as $U_{\rm eff}(\vec r_i)$. Then the Hamiltonian for edge 
particles can be written as follows,
\begin{eqnarray}
H_{edge}&=&\sum_{i=1}^{N_e}\biggl[(\frac{-i}{r_i}\frac{\partial}{\partial 
\phi_i}+
      \frac{\partial E_i^e}{\partial r_i})^2\label{Hedge}  \\ \nonumber&+& 
       (-i\frac{\partial}{\partial r_i}-
      \frac{1}{r_i}\frac{\partial E_i^e}{\partial \phi_i})^2
       -\frac{1}{r_i}\frac{\partial}{\partial r_i}
       +\frac{i}{r_i^2}\frac{\partial E_i^e}{\partial 
\phi_i}\biggr]\\ \nonumber
&+&\sum_{i<j}V_{sc}(\vec{r}_i-\vec{r}_j)+
\sum_i U_{\rm eff}(\vec{r}_i).    
\end{eqnarray}
By applying similar procedure to the edge Hamiltonian $H_{edge}$ , we arrive 
at the same Hamiltonian as $H'$ except that only edge electrons are 
included and $U$, $V$ are replaced by their modified forms.

Under the condition $\delta r_i<<R $, the transformed Hamiltonian can be 
reduced to the following one dimensional form,   
\begin{eqnarray}
H''_{1d} &=& 
-\frac{1}{R^2}\sum_{i}^{}\frac{\partial^2}{\partial\phi_i
^2}\\ \nonumber
&-&\frac{m}{R^2}\sum_{i<j}^{}\cot\frac{\phi_i-\phi_j}{2}(\frac{\partial}
{\partial\phi_i}-
\frac{\partial}{\partial\phi_j}) \\ \nonumber
&-&2K_0\sum_{i}^{}(\frac{-i}{R} 
{\frac{\partial}{\partial\phi_i}})+V_{sc}+U_{eff}
\end{eqnarray}
by making the substitutions $ z_i=Re^{i\phi_i} $
where  
\begin{equation}
K_0=\frac{|B^*|R}{2}-\frac{(N_e-1)m}{2R}.  
\label{k0}
\end{equation}
Then after a simple gauge transformation, which is equivalent to a translation 
of momentum along the edge that reads 
\begin{equation}
-i\frac{\partial}{\partial\phi_i} 
\rightarrow
-i\frac{\partial}{\partial\phi_i}+K_0,
\end{equation}
the linear term is cancelled, and  we get
\begin{eqnarray}
H''_{1d}&=& 
-\frac{1}{R^2}\sum_{i}^{}\frac{\partial^2}{\partial\phi_i
^2}-\frac{m}{R^2}\sum_{i<j}\cot\frac{\phi_i-\phi_j}{2}\\ \nonumber
&\times&(\frac{\partial}
{\partial\phi_i}-
\frac{\partial}{\partial\phi_j})
+V_{sc}+U_{eff}
\end{eqnarray}

We argue that the above Hamiltonian is in fact equivalent to the 
corresponding one describing the excitations of CSM if we switch off $V_{sc}$
and $U_{eff}$. To see it clearly, we can follow
a similar procedure to take a transformation of the CSM Hamiltonian.
\begin{eqnarray}
H_{cs}&=&\sum_i-\frac{\partial^2}{\partial 
\phi_i^2}+\frac{m(m-1)}{4}\sum_{i<j}
\biggl[\sin(\frac{\phi_{ij}}{2})\biggr]^{-2}
\label{HCS}
\end{eqnarray}
We separate the total wave function to a product of its ground state 
wave function $\Psi_{csg}$ and an extra part $\Phi'$ which contains 
all the information of the excitations of CSM,
\begin{equation}
\Psi_{cs}=\Psi_{csg} \Phi'
\end{equation}

Therefore we get an effective Hamiltonian with the same form as 
$H''_{1d}$ if we set  $R=1$,
\begin{eqnarray}
H''_{cs} &=&\Psi_{csg}^{-1}H_{cs}\Psi_{csg}\\ \nonumber
                      &=&-\sum_{i}^{}\frac{\partial^2}{\partial\phi_i
^2}-m\sum_{i<j}^{}\cot\frac{\phi_i-\phi_j}{2}(\frac{\partial}
{\partial\phi_i}-\frac{\partial}{\partial\phi_j})
\end{eqnarray} 

Moreover if we set the ground state wave function of $H_c$ to its 
one-dimensional limit, we arrive at exactly the ground state wave function of 
CSM.
\begin{equation}
\Psi_{csg}(\phi_1,...\phi_{N_e})=\prod_{i<j}\biggl[\sin\frac{\phi 
_{ij}}{2}\biggr]^m \\ \nonumber
\end{equation}      
From the above derivation one can see that the $\nu=1/m$ FQHE edge states
are described by the CSM if we switch off all the interactions between CFs.
Since the CSM is exactly solvable and its excitation wave functions can be 
calculated by using Jack's polynomials \cite {EX}, the study of FQHE edge 
states will pose no difficulty to us if we can properly handle the influence
of the interactions between edge particles.  
It will be discussed in the coming sections on what roles the interactions 
 play in the description of FQHE edge states .
One can see that the CSM has two-branches of the gapless excitations while 
the edge excitations of FQHE are chiral. This point will be clarified in  
Section IV.

\section {Non-renormalization of the exponents}

It is well-known that the CSM is an example of one dimensional ideal excluson 
gas(IEG) \cite {Wu} with the statistical parameter $m$. 
And the IEG is proved to describe the fixed point of the Luttinger
liquid. The bosonization of CSM shows that the low-lying excitations are 
governed by a $c=1$ CFT with the compactified radius $1/\sqrt{m}$
\cite{KY,Wu}. In this section, we would like to show that at least some kinds of 
short-range 
interactions between the CFs do not renormalize the topological exponent $g=m$
under the condition that the scatterings with large momentum transfer (including
 backward scattering and umklapp scattering) are absent because of the 
chirality.

To deal with the CSM with interactions,
we begin with the asymptotic Bethe ansatz (ABA) equation \cite{CS},
\begin{equation}
k_iL=2\pi I_i+\sum_{j}{'}\theta(k_i-k_j),
\label{ABA}
\end{equation}
where $k_i$ is the pseudomomentum of particle $i$, $L$ is the size of the one 
dimensional system concerned, and $I_i$ gives the corresponding quantum 
number, which is an integer or half-odd.
$\theta(k)$ represents the phase shift of a particle after a single collision 
with a pseudomomentum transfer of $k$.
It has been proved that ABA equations give exact solutions to the  energy 
spectrum of the CSM. We assume this approach could be generalized to the 
situations of CSM plus some other kind of interaction with force range 
shorter than $1\over r^2$ potential in the sense of perturbation. 
This assumption is justified for the 
following reasons: First, at the edge of fractional quantum 
Hall liquid with $\nu=1/m$, the 
linear density of the edge particles can be  estimated as
\begin{equation}
\rho\propto n\times l_B\propto B^{-1/2},
\end{equation}
where $n$ is the average bulk density of the FQH liquid that is fixed and $l_B 
=eB/m^*c$ is the magnetic length corresponding to the magnetic field $B$.
Under the condition of strong enough magnetic field, the edge particles can be 
regarded as a dilute one dimensional gas, where only two body collisions are 
important, and the free length between two collisions is long enough to allow 
the phase shift to reach its asymptotic value. Secondly, what we are concerned
with is the property of low energy excitations near the Fermi surface, not 
the whole precise energy spectrum which can not be given by ABA. 
The low energy excitations involve only scattering processes with small 
momentum transfer $\Delta k$ (because of Chirality, see Sec.IV), and are 
determined by the behavior
of $\theta(k) $ around $k=0$, which is dominated by $\theta_{cs}(k)$
(see below).
Under the condition of low energy limit where we let $\Delta k$ approach 
zero slowly, $\theta(k)$ will become asymptotically close to $\theta_{cs}(k)$,
as a result we can expect ABA calculations to give asymptotically correct 
results. Here we implicitly assume : the low energy spectrum of the CSM varies
continuously with respect to the addition of small perturbation without 
undertaking any abrupt changes like a phase transition. This assumption is 
reasonable, for the low energy limit of the CSM is the fixed point of 
Luttinger liquid which is robust against perturbations. Therefore, what 
follows from ABA, as we believe, is credible.     
 As a matter of fact, for a large class of short-range interactions, we can 
expect the ABA works in describing the low-lying excitations of the system.
Indeed, there are several kinds of short-range interactions whose low-lying
excitations are governed by the ABA. An example of them is the 
$\delta^{(l)}$-function 
interaction with $(l)$ representing the l-th derivative of the $\delta$-function 
and
$l$ being restricted to $l<m$. The pseduopotentials used by Haldane 
\cite{psudo} are other examples because of the vanishing of the expectation 
value of the pseudopotentials in the ground state.
 
To calculate the phase shift, we note the  analog of the Schrodinger 
equation of the two-body CSM with an additional short-range interaction in the 
limit 
$L\to \infty$ to the radial
equation of a three-dimensional scattering problem of a centrally
symmetric potential. The topological exponent $m$ corresponds to the 
total angular momentum $l$, i. e. $m=l+1$. The Schrodinger equation reads
\begin{equation}
\frac{d^2\psi(x)}{dx^2}+\biggl[(E-V)-\frac{l(l+1)}{x^2}\biggr]\psi(x)=0.
\label{SEC}
\end{equation}
The asymptotic solution of (\ref{SEC}) for $x\gg 0$ is given by
\begin{equation}
\psi(x)\approx 2\sin(kx-\frac{1}{2}l\pi+\delta_l),
\end{equation}
where $\delta_l$ is the three-dimensional phase shift corresponding to the
scattering potential $V$. In the sense of 1-d scattering,
\begin{equation}
\theta(k)=\pi(m-1){\rm sgn}(k)-2\delta_l(k).
\label{ps}
\end{equation}
We see that the contribution of $V$ to the phase shift is
\begin{equation}
\theta_{\rm reg}(k)=-2\delta_l(k),
\end{equation}
 which is continuous and vanishing at $k=0$ if $V$ is short-ranged (shorter
 than $1/r^2$).

Now, let's make the relation to the macroscopic theory. In terms of the 
partition function (\ref{apf}), there is a most probable edge CF number
$\bar{N}^e$ which is given by $\delta Z/\delta N^e=0$.
$\bar{N}^e=\int dx \rho(x)$ with the edge density
$\rho(x)=h(x)\rho_e$ \cite{Wen}. Here $h(x)$ is the edge deformation and 
$\rho_e$ is the average density of the bulk electrons.  We do not distinguish
$\bar{N_e}$ and $N_e$ hereforth if there is no ambiguity.
The low energy properties of the CSM can be obtained from the  
ABA equations, 
\begin{eqnarray}
    \rho(k)&=&\rho_0(k)-\int\limits_{-k_F}^{k_F}g(k-q)\rho(q) dq \\ 
    \epsilon(k)&=&\epsilon_0(k)-\int\limits_{-k_F}^{k_F}g(k-q)\epsilon(q) dq 
\label{tba}
\end{eqnarray}
where $k_F=\pi m N^e_0/L$,
\begin{equation}
    g(k)=\frac{1}{2\pi} \frac{d\theta(k)}{dk},
\end{equation}
$\epsilon_0(k)=k^2-k_F^2$ and  $ \rho_0(k)=\frac{1}{2\pi}  $.

If we consider only the CSM without other interactions ,then we have
\begin{equation}
\theta_{cs}(k)=\pi(m-1){\rm sgn}(k).
\label{theta} 
\end{equation}
Substituting (\ref{theta}) into (\ref{tba}) and after the linearization, we get
\begin{eqnarray}
\epsilon_{cs\pm}(k)=\biggl\{{\begin{array}{ll}\pm v_+(k\mp k_F),
 &{\rm if}~ |k|>k_F \\   

 \pm v_-(k\mp k_F),&{\rm if}~ |k|<k_F,\\ \end{array}}   
  \label{LD}   
\end{eqnarray}
where
\begin{eqnarray}
v_+&=&\frac{d\epsilon(k)}{dk}\biggl {|}_{k=k_F+0^+}=v_F \\ \nonumber
v_-&=&\frac{d\epsilon(k)}{dk}\biggl {|}_{k=k_F-0^+}=\frac{v_F}{m}
\end{eqnarray}
with $   v_F=2k_F  $ 
and
\begin{eqnarray}   
      \rho_+&=&\rho(k_F+0^+)=\frac{L}{2\pi}  \\ \nonumber
      \rho_-&=&\rho(k_F-0^+)=\frac{L}{2\pi m} 
\end{eqnarray}      
We rewrite the important equations essential to the bosonization for CSM as 
follows
\begin{eqnarray}
       v_+&=&mv_-  \\
       \rho_+&=&m \rho_-.
\label{vrho}
\end{eqnarray}

A successful bosonization of the theory with the refraction dispersion 
(\ref{LD}) has been done by the authors of \cite{Wu} and one shows that the
low-lying excitations of the CSM are controlled by the $c=1$ CFT with its 
compactified radius ${\cal R}=1/\sqrt{m}$ \cite{KY}. This implies that the 
low-lying 
states
of the CSM have the Luttinger liquid behaviors with the exponent $g=m$. 
We will be back to this issue later after we supplies the chiral constraint 
and then show that the edge states of FQHE have the CLL behaviors. 

Now, let us see the effects of the interactions.
Following our discussion that leads to the $c=1$ CFT with the compactified 
radius ${\cal R}=1/\sqrt{m}$, the relations (\ref{vrho}) are essential. 
We would 
like to check if they are renormalized by the interactions between CFs.
Here, we limit our discussion to the case $m\not=1$. We assume the ABA works 
to describe the low-lying excitations of the system with an additional short 
range interaction, which is consitent with the chirality of the edge 
excitations.      
Differentiating the phase shift (\ref{ps}) with respect to $k$, one has
\begin{equation}
g(k)=(m-1)\delta(k)+g_{\rm reg}(k).
\end{equation}
   
The continuity of $\theta_{\rm reg}$ implies that $g_{\rm reg}$ is no
more singular than the $\delta$-function at $k\to 0$. Therefore, we 
can prove that the relations 
(\ref{vrho}) still hold even after we have introduced a short-range 
interaction. After differentiating the dressed energy equation 
(\ref{tba}) that is assumed holding for the short-range interaction we are
using and in the dilute gas approximation,  with respect to k, 
we obtain
\begin{eqnarray}
 v_\pm&=&v_0+\int\limits_{-k_F}^{k_F}\epsilon(q)\frac{d}{dq}g(k_F\pm 0^+-q) 
dq\\ \nonumber
      &=&v_0-\int\limits_{-k_F}^{k_F}\frac{d}{dq}\epsilon(q)g(k_F\pm 0^+-q) 
dq\\ \nonumber
      &+&\epsilon(k_F)g(k_F\pm 0^+-k_F) \\ \nonumber
       &-&\epsilon(-k_F)g(k_F\pm 0^++k_F). 
\end{eqnarray}
The definition of $k_F$ ,i.e., $\epsilon(\pm k_F)=0$, leads to  
\begin{eqnarray}
 v_+-v_- & = & 
\int\limits_{-k_F}^{k_F}\frac{d}{dq}\epsilon(q)(m-1)\delta(q-k)\\ 
\nonumber
        & = & (m-1)v_- . 
\end{eqnarray}             
Hence 
\begin{equation}
v_+=m v_-.
\end{equation}
The value of $v_\pm$ can be modified by the interactions but the above relation
does not change. 
Note that if $\epsilon(k_F)=0$ and $\epsilon(-k_F)g(k_F\pm 0^++k_F)$ is 
continuous at $k_F$, the above conclusion still holds, which will be the
case in the CLL derivation of Sec IV.
By performing a similar procedure to $\rho(k) $,
we can show (\ref{vrho}) for $\rho_\pm$ as well. Therefore one can see that the 
bosonization
process of the CSM is still applicable in the presence of perturbative 
interactions,  
and the topological exponent $g=m$ is not renormalized by the short-range 
interaction.
As a result, the compactified radius of the $c=1$ CFT which governs the 
low-lying
excitations of the theory does not change.

Let us give more comments on the conclusion drawn above. This result seems 
remarkable at 
the first sight, when compared with the standard Luttinger liquid theory, in 
which we will have the characteristic exponent renormalized once a short-range
perturbative interaction is switched
on. In fact, no inconsistences exist here. In the bosonization of the general 
Luttinger liquid, only short
range interactions are considered, whose Fourier transformation $V(k)$ at k=0 
possesses no singularity.
Even if the divergence of $ V(k)$ as $k$ approaches zero does show up, it is 
suppressed by introducing something like
a short-range cutoff or a long-range cutoff which makes the problem concerned 
more subtle. The exponent
so obtained may be cutoff-dependent. So we can not naively apply it here. In 
contrast to the standard approach, the bosonization of the CSM is based
on the especially simple form of the phase shift function of the $1/r^2$ 
interaction 
that is essential to the solution of ABA.
The singularity here manifests itself as a step discontinuity which can be 
handled easily (no cutoff is needed). Because of the 
critical property of the $1/r^2$ interaction, no other interactions with 
shorter ranges can alter this discontinuity,
which guarantees the robustness of the bosonization process. In short, the 
bosonization of CSM is not so general as the standard one, but it surely 
makes a step forward in understanding the low energy physics of nontrivial 
interactions.  
    
We emphasize once again that both $1/r^2$ interaction and the 
chirality contribute to the robustness of $g=1/m$ when 
$m>1$. In general, the critical exponent will be changed by the introduction of
other short-range interactions if the chirality is not present and backward 
scattering
is allowed. In contrast, in case of $m=1$, where we are actually 
dealing with a Fermi liquid, the discontinuity of the phase shift $\theta(k)$ 
is absent. So the above argument of robustness fails. 
An simple example is to consider a $\delta$-function interaction. For $m>1$,
the short-range divergence of the $1/r^2$ potential requires that the wave 
function
vanishes when two particles approach each other. Hence  
the $\delta$-function contribution to the phase shift is completely 
suppressed in case of $m>1$,while it does show up for
 $m=1$ \cite{YY}. On the occasion of $m=1$, however, the 
chirality alone serves as the determinant factor to ensure 
the non-renormalizability of $g=1$, by prohibiting the left-right scattering 
part of perturbative interactions from modifying $g$ . Therefore, one can see 
that 
the 
different microscopic mechanisms for $m>1$ and $m=1$ give the same 
macroscopic result.

From the above arguments, we see that the topological exponent is invariant to
the perturbations introduced by additional interactions between particles,
if their interaction range is shorter than that of $ 1/r^2$.
However, the long range nature of Coulomb interaction allows it
to dominate the $1/r^2 $  interaction which gives $g=\nu$. Considering its 
especially singular behavior at k=0, we believe that the so called topological
index can no longer survive, if an unscreened Coulomb interaction without
any cutoff really exists. Fortunately, we have several possibilities that will
lead to  partial screening of the Coulomb interaction. In real experiments,
the edge electrons actually are not isolated to a  wire-like structure.  
There are bulk electrons adjacent to them, which can provide mirror
charges and reduce the original Coulomb interaction
to a shorter range interaction. What is more, metal electrodes commonly
used in experiments to supply a confinement potential can also serve as 
a mirror charges provider. So we only have to concern ourselves with partly 
screened or drastically renormalized Coulomb
interaction instead of the bare one. The effect of short-range interactions has 
been 
discussed in this section. We will consider the case where the 
Coulomb interaction is drastically renormalized later after we explain the
chirality of the edge states.

\section{ chiral Luttinger liquid: the microscopic point of view}

\subsection{Microscopic Derivation of CLL from the Radial Equation}

In the previous two sections, we freeze the radial degree of freedom of
the edge particles and see that the azimuthal dynamics is described by the CSM. 
However, there are two branches of gapless excitations in the CSM and the 
chirality of the edge excitations are not shown. To arrive at the conclusion of 
chirality, we 
take the radial degree of freedom into account. Let us first make some 
simplifications before
going into details. First, the interactions between CFs are assumed
to be independent of the radial degree of freedom because of the small width of 
the edge. Moreover,
we can think of the interaction between the CFs as consisting
 of only the $1/x^2$-type since
we have demonstrated that short-range interactions do not renormalize the 
topological exponent $g=m$. At the end of this section, we will give a 
discussion 
about the effect of Coulomb interaction when it is drastically renormalized.
Secondly, the  edge potential is assumed as follows,
\begin{eqnarray}
U_{\rm eff}(r)=\biggl\{{\begin{array}{lll}
               \infty,& {\rm if }~r\geq R, \\ 

               eE(r-R'), & {\rm if}~R'<r<R, \\ 
               0,& {\rm if }~r<R'
               \end{array}}
\end{eqnarray}
Here $E$ is the value of the external field and  
one takes $v_d^*=cE/|B^*|>R_c^{*}\omega^*_c$. The exact value of 
$R'$ is not important. The only requirement is $R'\simeq 
R-R_c^*$.  In Sec. II, we take the approximation in the edge Hamiltonian
(\ref{Hedge}) with $r_i\simeq R$ and arrive at the CSM. To consider the radial 
degree 
of freedom, we separate the wave function into the azumithal 
and radial parts near the edge,
\begin{equation}
\Psi_{c}(z_1,...,z_{N_e})=g(r_1,...,r_{N_e})\Psi_{cs}
(\varphi_1,...,\varphi_{N_e}),
\end{equation}
where $\Psi_{cs}$ is the eigenstate wave function of the CSM, which is 
antisymmetric and $g$ is a symmetric radial wave function. 
Substituting the separated form of the wave function to the 
Schrodinger equation corresponding to $H_{edge}$ 
(\ref{Hedge}), one finds that the equation can be reduced to the 
radial eigenstate equation which reads
\begin{eqnarray}
&&\sum_i\biggl[-\frac{\partial^2}{\partial r_i^2}+
(\frac{n'}{r_i}-\frac{|B^*|}{2}r_i)^2 \biggr]g(r_1,...,r_{N_e})\\ 
\nonumber
&&+[U_{eff}+O(\delta r_i/R)]g(r_1,...,r_{N_e})=Eg(r_1,...,r_{N_e}).
\end{eqnarray}
where $ n'=n+{m\over 2}(N_e-1)$ and the terms 
$\frac{1}{R}\frac{\partial}{\partial r_i}$ have been 
absorbed into $g$ by a simple transformation like the multiplication of $e^{\sum 
r_i/R}$.
One can see that the radial eigenstate equation can be 
treated in the single particle picture except that the pseduomomenta $k=n R$
are related to one another by the ABA equations(\ref{ABA}). It is reasonable to 
arrive 
at such a result because the interactions between CFs are the functions of 
$\vec{r_i}
-\vec{r_j}$ and  the radius-dependent part of the interactions is of order
$\delta r/R$. Now we employ the harmonic approximation
used by Halperin in the case of IQHE edge states\cite{Halp}.
Let us first turn off the applied electric field. The radial single particle
wave equation in the stripe approximation reads
\begin{equation}
-\frac{d^2g}{dy^2} +B^{*2}y^2g=\varepsilon_+g,
\label{+}
\end{equation} 
for $ n'>0$ and
\begin{equation} 
-\frac{d^2g}{d y^2} 
+B^{*2}y^2g+|n' B^*|g=\varepsilon_- g,
\label{-}
\end{equation}    
for $ n'<0$.       
Here $y=r-R_{n'}$ and 
\begin{equation}
R_{n'}=\sqrt{\frac{2|n'|}{|B^*|}}.
\end{equation}
Comparing (\ref{+}) with (\ref{-}), we see that the magnetic field separates the
$n'<0$ sector from the $n'>0$ sector by an energy gap $|n'|\hbar\omega_c^*$.
Therefore only the $n'>0$ (or equivalently, $ k>-k_F$) sector needs to be 
considered 
for the low-lying excitations. This is the first sign of chirality. 
The harmonic equation (\ref{+}) has its eigenstate energy  
\begin{equation}
\varepsilon_{+,\nu^*}=\hbar\omega^*_c((\nu^*-1)+\frac{1}{2}),
\end{equation}
if the center of the harmonic potential $R_{n'}\ll R$. This is consistent with 
the mean-field
approximation to the bulk state because $R_{n'}\ll R$  actually corresponds to
the bulk state of the theory if we recognize that the width of the harmonic
oscillator wave function is about several times  the cyclotron motion radius 
$R_c^*$. Since $R_{n'}$ is the function of $n'$, $p_n=m^*\omega^* R_{n'}$
can be regarded as a momentum-like quantity. The harmonic oscillator
energy for $R-R_{n'}\ll R_c^*$ implies that there is no left-side Fermi point.
This provides a necessary condition of the chirality. To justify the CLL,
one should show the existence of gapless excitations on the right side. It is 
known that the eigenstate energy at $R_{n'}=R$ is raised to 
\begin{equation}
\varepsilon_{R,\nu^*}=\hbar\omega^*_c\biggl(2(\nu^*-1)+\frac{3}{2}\biggr),
\end{equation}
because of the vanishing of the wave function at $r=R$. One asks that what
happens if $R_{n'}$ is slightly away from $R$. To see it clearly, we rewrite
(\ref{+}) as
\begin{eqnarray}
&&-\frac{d^2g}{d\tilde y^2} +B^{*2}\tilde y^2g
+2B^{*2}(R-R_{n'})\tilde y g\nonumber \\ &&+
B^{*2}(R_{n'}-R)^2 g=\varepsilon_+g, \label{++}
\end{eqnarray} 
where $\tilde y=r-R$. If $R_{n'}$ is very close to $R$, i. e., $|R-R_{n'}| 
\leq R_c^*$, one can take the 
third term as perturbation and a first order perturbative calculation shows
\begin{equation}
\delta\varepsilon_{0,\nu^*}=\varepsilon_{+,\nu^*}-\varepsilon_{R,\nu^*}
=v_c^*(p_n-p_R)+(p_n-p_R)^2,
\label{e0}
\end{equation}
where $v_c^*=2\pi^{-1/2}l_{B^*}\omega^*_c$ is of the order of the cyclotron
velocity of the CF corresponding to $B^*$ and then of the order of $v_F$.
So, if we take $v_c^*\approx v_F$ and note that $p_n-p_R\approx
k-K_0$, the dispersion (\ref{e0}) can be simply
rewritten as
\begin{equation}
\delta\varepsilon_0(k)\approx(k-K_0+k_F)^2-k_F^2,
\label{dr}
\end{equation}
where $k$ is given by the ABA equations (\ref{ABA}). Near the Fermi point
$K_0$, the dispersion can be linearized as
\begin{equation}
\delta\varepsilon_0(k)=v_F(k-K_0),
\end{equation}
which implies that there is a right-moving sound wave excitation  along the
edge with the sound velocity $v_F$.
There is another Fermi point $k=K_0-2k_F$, which corresponds to 
$R_{n'}\approx R-2R_c^*$ and is outside of the region we are considering 
and in fact belongs to the bulk state. 
To show that the above chiral theory has a Luttinger liquid 
behavior, we extend continuously $\delta\varepsilon_0(k)$ to all 
possible pseudomomenta which obey the ABA (\ref{ABA}). Then,
the equation (\ref{dr}) and (\ref{ABA}) mean that the system
is an IEG \cite{Wu}. The problem can be solved by 
using a bosonization procedure developed in ref. \cite{Wu}.  
The edge excitations can be obtained by considering only the properties of 
such an IEG system near $k\sim K_0$. Consequently, 
the low-lying excitations of the theory are controlled by the $c=1$ CFT with 
its compactified radius ${\cal R}=1/\sqrt{m}$ as we point out in 
the discussion of the CSM in Sec. III. However, the relevant excitations of 
the edge states include only the right-moving branch. In  
other words, the edge states are chiral and the sound wave excitations  
correspond to the non-zero modes of the right-moving sector of 
the $c=1$ CFT. There are two other kinds of edge excitations which  
correspond to the particle additions to the
ground state and the current excitations along the edge respectively. The 
velocity 
relations of these excitations are given by \cite{Wu}
\begin{equation}
v_M=mv_F, v_J=v_F/m, v_F=\sqrt{v_Mv_J}.
\label{vr}
\end{equation}
The relations resemble those of Haldane's 
Luttinger liquid if one identifies $m$ with the characteristic
parameter $e^{-2\varphi}$ in the Luttinger liquid theory \cite{Hald}. These 
observations are
crucial to the conclusion that the edge states are controlled by the $c=1$ CFT 
with its 
compactified radius ${\cal R}=1/\sqrt{m}$. 

To arrive at the effective theory of CLL, let us perform the following 
bosonization procedure.

 According to 
(\ref{dr}) and (\ref{ABA}),  the edge excitations with the pseudomomentum $k$ 
have their dressed energy 
\begin{eqnarray}
\varepsilon(k)=\biggl\{{\begin{array}{ll}
  (k^2-k^2_F)/m, &|k|<k_F, \\ 
   k^2-k^2_F,&|k|>k_F. \\ \end{array}}
\end{eqnarray}
Here we have made a translation $k\to k+K_0-k_F$. The linearization 
approximation of 
the dressed energy near 
$k\sim\pm k_F$ is given by (\ref{LD}).
In terms of the linearized dressed energy, we obtain a free fermion-like
representation of the theory and then can easily bosonize it 
\cite{Wu}. The Fourier transformation of the right-moving density operator is
given by
\begin{eqnarray}
\rho_q^{(+)}&=&\displaystyle \sum_{k>k_F}:c^\dagger_{k-q}c_k:
+ \displaystyle \sum_{k<k_F-m q}
:c^\dagger_{k+mq}c_k:\\
& &+ \displaystyle
\sum_{k_F-m q< k < k_F}:c^\dagger_{\frac{k-k_F}
{m}+k_F+q}c_k:,
\end{eqnarray}
for $q>0$ is the sound wave vector. Here $c_k$ is a fermion annihilation 
operator. And a similar $\rho_q^{(-)}$ can be defined near $k\sim -k_F$. 
The bosonized Hamiltonian is given by
\begin{equation}
H_B=v_F\{ \sum_{q>0}q(b_q^\dagger b_q
+\tilde{b}_q^\dagger \tilde{b}_q)
+\frac{1}{2}\frac{\pi}{L}[m M^2
+\frac{1}{m} J^2] \}.
\label{bosonH}
\end{equation}
Thus, we have a current algebra like
\begin{equation}
[\rho_q^{(\pm)},\rho_q^{(\pm)\dagger}]=
\frac{L}{2\pi} q\, \delta_{q,q'},~~
[H_B,\rho_q^{(\pm)}]= \pm v_Fq\rho_q^{(\pm)}.
\label{density}
\end{equation}

In the coordinate-space formulation, the normalized
density field $\rho(x)$ is given
by $\rho(x)=\rho_R(x)+\rho_L(x)$:
\begin{equation}
\rho_L(x)=\frac{M}{2L}+
\displaystyle\sum_{q>0}\sqrt{q/2\pi Lm}
(e^{iqx}b_q+e^{-iqx}b_q^\dagger),
\label{rhofield}
\end{equation}
and $\rho_R(x)$ is similarly constructed from
$\tilde{b}_q$ and $\tilde{b}_q^\dagger$. 
Here ${b}_q=\sqrt{2\pi/qL}{\rho}_q^{(+)\dagger}$  and so on. The boson
field $\phi(x)$, which is conjugated to $\rho(x)$
and satisfies $[\phi(x),\rho(x')]=i\delta(x-x')$,
is $\phi(x)=\phi_R(x)+\phi_L(x)$ with
\begin{displaymath}
\phi_L(x)= \frac{\phi_{0,L}}{2}+\frac{\pi Jx}{2L}+i
\displaystyle \sum_{q>0}\sqrt{\pi m/
2qL}(e^{iqx}b_q-e^{-iqx}b_q^\dagger),
\end{displaymath}
and a similar $\phi_{L}(x)$. Here $M$ and $J$ are operators
with integer eigenvalues, and $\phi_0=\phi_{l0}+\phi_{r0}$ is an angular 
variable conjugated to M: $[\phi_0,M]=i$. The Hamiltonian (\ref{bosonH})
becomes
\begin{equation}
H_B =\frac{v_F}{2\pi}
\int_0^Ldx\; [\Pi(x)^2+(\partial_xX(x))^2],
\label{fieldH}
\end{equation}
where $\Pi(x)=\pi m^{1/2}\rho(x)$ and
$X(x)=m^{-1/2}\phi(x)$. With
$X(x,t)=e^{iHt}X(x)e^{-iHt}$, the
Lagrangian density reads
\begin{equation}
{\cal L}=\frac{v_F}{2\pi}\,\partial_\alpha
X(x,t)\,\partial^\alpha X(x,t).
\end{equation}
We recognize that ${\cal L}$ is the Lagrangian
of a $c=1$ CFT. Since $\phi_{0}$
is an angular variable, there is a hidden
invariance in the theory under $\phi\to\phi+2\pi$.
The field $X$ is thus said to be ``compactified''
on a circle, with a radius that is determined
by the exclusion statistics:
\begin{equation}
X\sim X+2\pi {\cal R},~~~ {\cal R}^2=1/m.
\end{equation}
States $V[X]|0\rangle$ or operators
$V[X]$ are allowed only if they respect
this invariance, so quantum numbers of
quasiparticles are strongly constrained. 

In the present case, only the right-moving sector is relevant. 
So, we have an `almost' chiral edge state theory whose sound wave excitation is
chiral but there are charge leakages between the bulk and the edge. 
The leakages are reflected in the zero-mode particle number and current 
excitations 
\cite{foot}. In this almost chiral theory, the charge-one 
fermion operator is defined by
\begin{equation}
\Psi_R^\dagger(x)=\sum_{l=-\infty}^\infty\exp(i2(l+\frac{1}{2}m)\theta_R(x))
\exp(i\phi_R(x)),
\end{equation}
where 
\begin{equation}
\theta_R(x)=\pi\int_{-\infty}^x \rho_R(x')dx'.
\end{equation}
The correlation function, then, can be calculated \cite{EX}
\begin{eqnarray}
<\Psi_R^\dagger(x,t)\Psi_R(0,0)>&=&\sum_{l=-\infty}^{\infty}
C_l\biggl(\frac{1}{x-v_Ft})^{(l+m)^2/m} \nonumber\\ 
&&\exp(i(2\pi(l+\frac{1}{2})x/L)).
\end{eqnarray}
The $l=0$ sector recovers Wen's result \cite{Wen}.
In other words, the present theory justifies microscopically Wen's 
suggestion of CLL of the FQHE edge states . 

Now, let us turn on the electric field. The external scalar potential 
perturbation 
is added to the harmonic oscillator equation
(\ref{++})  
\begin{eqnarray}
&&-\frac{d^2g}{d\tilde y^2} +B^{*2}\tilde y^2g 
+2B^{*2}(R-R_{n'})\tilde y g \nonumber\\ &&+ 
B^{*2}(R_{n'}-R)^2 g+eE(r-R')g=\varepsilon_+g.
\label{+e}
\end{eqnarray} 
Up to a zero position shift,  (\ref{+e}) yields a harmonic potential
centered at $R$ perturbed by 
\begin{equation}
(2B^{*2}(R-R_{n'})+eE)\tilde y.
\end{equation}
A perturbative calculation to the second order shows 
\begin{equation}
\delta\varepsilon_{d,0}(k)\approx(k-K_0+k^*_F)^2-k_F^{*2},
\label{drd}
\end{equation}
where $k_F^*=\frac{1}{2}v_d^*+k_F$ with $v_d^*=cE/|B^*|$, the drift velocity
of a CF in the effective field $B^*$. So, (\ref{drd}) and
(\ref{ABA}) define an new IEG system. All discussions made above for the 
situation
with no electric field are valid after a replacement
of $k_F$ by $k_F^*$. One notes that $v_F\ll v_d^*$. So, $v_F^*=2k_F^*
\approx v_d^*$. The current velocity $v_J=v_d$ is related to the sound wave
velocity $v_s\approx v_d^*$ by Haldane's velocity relation $v_J=v_s/m$.

\subsection{ Further Discussion of Coulomb Interaction}

In this subsection, we give a further discussion of the effects of the 
interactions between CFs
. The short-range interactions do not renormalize the topological
exponent $g=m$ because the conclusion drawn from Sec. III is also applicable
to the last subsection. Here what interests us is the Coulomb 
interaction.

The wave equation of the one-dimensional Hamiltonian (\ref{HCS}) with an 
additional Coulomb interaction (4.29)
for $N_e=2$ can be exactly solved  in the large $L$ limit.
\begin{equation}
V_{\rm coul}=\frac{\alpha\pi}{L|\sin(\pi x_{12}/L)|}
\end{equation}

Actually, this 1d Hamiltonian is the same as the radial part of the Hamiltonian 
of 
an electron scattered by a three-dimensional Coulomb potential
\cite{LL}. The phase shift, is then given by
\begin{equation} 
\theta(k)={\rm sgn}(k)\tilde\phi\pi-2{\rm arg}\Gamma(m+i/k).
\end{equation}
By going to the dilute gas limit, $x_1\ll x_2\ll...\ll x_{N_e}$, the $N_e$ body
problem can be solved asymptotically by means of the Bethe ansatz with the trial 
wave 
function
\begin{eqnarray}
&&\Phi_s(x_1,..,x_{N_e})=\sum_P A(P)\exp\{i\sum_ik_{P_i}x_i\\ \nonumber
&&-i\sum_{i<j}F(k_{P_i}-k_{P_j}, x_i-x_j)\}, \\ \nonumber
&&F(k,x)=\frac{\alpha}{2k}\ln(\biggl[\tan\frac{\pi x}{2L}\biggr]\frac{2\pi k}
{L}).
\end{eqnarray}
The asymptotic Bethe ansatz equations are given by
$
n_i=I_i+\frac{1}{2\pi}\sum_{j\not=i} \theta(k_i-k_j).
$
Then we assume the Coulomb interaction to be drastically renormalized so 
that $\alpha L\ll 1$, which means that the minimal pseudomomenta spacing
$\delta k\gg \alpha$. In the thermodynamic limit at low temperature,
the thermodynamic Bethe ansatz equation for $\rho(k)$ can be solved 
iteratively. By integrating $\rho(k)$ with respect to $k$, one has  
\begin{equation}
k=\tilde k+C\ln(k-k_F),
\end{equation}
where $\tilde k$ is regular when $k-k_F$ tends to zero and  $C$ is a constant in
proportion to $\alpha L$. Hence, the radial equation reads
\begin{eqnarray}
&&-\frac{d^2 g}{dy^2}+\omega^{*2}_c y^2 g\\ \nonumber
&&+C'y\ln(p_n-p_R) g
=\varepsilon g,
\end{eqnarray}
where $y=r-R_{\tilde n}$. By taking the 
approximation $r\simeq R$ in the logarithmic term, one can see that the 
dispersion relation reads, after switching on the external electric field,
\begin{equation}
\delta\varepsilon_{\tilde n,\nu^*}=(p_n-p_R)(v_d^*+A\ln(p_n-p_R)),
\end{equation}
with $A$ proportional to $\alpha L$.
We then obtain a branch of excitations with the dispersion $q\ln q$, which is
precisely the 1d plasmon excitation caused by the Coulomb interaction 
\cite{Wen1}.   

\section{edge states with other odd denominator filling factors}

In the rest of the paper, we would like to generalize the approach
developed in the previous sections to deal with the other odd denominator
filling factors. In general, one can transform the electronic Hamiltonian to 
the CF Hamiltonian like
\begin{eqnarray}
H_{\rm cf}&=&\sum_{I=1}^P\sum_{i_I=1}^{N_I}\frac{1}{2m^*}\biggl[-i\hbar
\frac{\partial}{\partial \vec{r}_{i_I}}+\frac{e}{c}\vec{A}_{i_I}\\ \nonumber
&+&\frac{e}{c}\vec{a}_{i_I}\biggr]^2+\sum_{I}(V_I+U_I),
\end{eqnarray}
where $I=1,...,P$ are the CF Landau level indices and $\sum N_I=N_e$. 
The statistic gauge field is defined by
\begin{equation}
\vec{a}_{i_I}=\phi_0\sum_{J=1}\sum_{j_J}{'}K_{IJ}\frac{\hat{z}\times
(\vec{r}_{i_I}-\vec{r}_{j_J})}{| \vec{r}_{i_I}-\vec{r}_{j_J})|^2},
\end{equation}
where the prime symbol means that $j_J\not= i_I$ if
$I=J$. The matrix $K_{IJ}$ is symmetric with odd diagonal elements, which 
is introduced by Wen and Zee \cite{WZ} to describe the effective field theory of
the FQHE. By using a gauge invariance argument, Wen has pointed out that a
positive eigenvalue corresponds to a branch of right-moving edge excitations
while a negative one to a left-moving branch \cite{Wen}. Here, we would like 
to give a microscopic justification of the macroscopic effective theory of the
edge states.
Applying the mean-field approximation to the bulk states, one can see that
\begin{equation}
\nu=\sum_{IJ}(K^{-1})_{IJ}.
\end{equation} 
For explicity, we consider the case of multi-layer FQHE. In this case, the 
electronic ground state wave function can be simply written out in the 
lowest Landau level version,
\begin{eqnarray}
\Psi_e(z_1,...,z_{N_e})&=&\prod_{I=1}^P\prod_{i_I<j_I}(z_{i_I}-z_{j_I})^{K_{II}}
\\ \nonumber
&&\prod_{I<J}^P\prod_{i_I,j_J}(z_{i_I}-z_{j_J})^{K_{IJ}}\\ \nonumber
&&\times\exp(-\frac{1}{4}
\sum_{I, i_I}|z_{i_I}|^2).
\end{eqnarray}
By conducting the transformations similar to those we use in the case of 
$\nu=1/m$, we 
arrive at a 1-d generalized CSM whose Hamiltonian reads,
\begin{eqnarray}
H_{gcs}&=&\sum_{I,i_I}-\frac{\partial^2}{\partial x_{i_I}^2}\nonumber \\
&+&\frac{\pi^2}{2L^2}
\sum_{I,J;i_I\not=j_J}g_{IJ}\biggl[\sin \frac{\pi(x_{i_I}-x_{j_J})}{L}\biggr]
^{-2}.
\end{eqnarray} 
The corresponding many-body problem is also exactly solvable and the 
corresponding
Bethe ansatz equations read as follows
\begin{equation}
Lk_{i_I}=2\pi I_{i_I}+\sum_J\sum_{j_J}{'}(K_{IJ}-1){\rm sgn}(k_{i_I}-k_{j_J}).
\label{mba}
\end{equation}
The radial wave equation for a single particle is given by 
\begin{equation}
-\frac{d^2g_I}{dy_I^2}+B^{*2}(\frac{n'_I}{r_I}-\frac{|B^*|}{2}r_I)^2g_I+
|B^*|
v^{*}_{Id}y_Ig_I=\varepsilon _I g_I,
\end{equation}
where $v^*_{Id}=cE_I/|B^*|$ and $E_I$ is the applied electric field 
corresponding to the potential $U_I(r_I)$. Then, a procedure parallel to
that discussed in the previous sections leads to $p$-branches
of edge excitations with sound wave velocities $v_{IF}\approx v^*_{Id}$ for 
the $I$-th branch. This 
recovers the CLL theory \cite{Wen}. Although our explicit derivation is applied 
to
the multi-layer quantum Hall liquid, we believe that the theory also holds 
in the single-layer case because the explicit ground state wave function is
not crucial. What is important is the Bethe ansatz equations (\ref{mba}), which
as we believe are valid for the single-layer case although the explicit 
derivation
needs more work. 

\section{CONCLUSIONS}

In conclusion, we have given the detailed derivation of a microscopic model of 
the 
CF-type excitations of the FQHE edge with odd denominator filling factor $\nu$. 
We start with the general Hamiltonian for a two-dimensional system of 
interacting electrons in a strong perpendicular magnetic field and confined to 
some specific geometry like a disc. The potential at the edge is 
assumed to maintain a sharp shape. After a CF type of anyon transformation, 
we obtain a Hamiltonian whose ground 
state wave function can be simply written out. By extracting the excitation 
part from the wave function, we arrive at a form of Hamiltonian which 
is shown to be equivalent to the corresponding Hamiltonian of the CSM when 
reduced 
to one dimension. We then take into account the influences of
the interactions between edge particles. We demonstrate that the 
bosonization process specific to CSM is not changed by the introduction 
of some kinds of short-range interactions. The characteristic exponent 
$g=1/m$ is indeed
robust against perturbations under the condition of chirality. The low-lying 
excitations of the CF system
are governed by a $c=1$ CFT with its compactified radius ${\cal R}=1/\sqrt{m}$
for the case of $\nu=1/m$. By taking the radial degree of freedom into account,
we show that the microscopic theory indeed justifies the macroscopic CLL.  
We also generalize the approach applied to $\nu=1/m$ to other cases with odd
denominator filling factors.

\bigskip

\noindent{\bf ACKNOWLEDGMENTS}

\bigskip

The authors are grateful to Z. B. Su  for helpful 
discussions and  would like to thank Y. S. Wu for his important comments 
in the earlier version of this paper and useful discussions.  One of the 
authors (W.J.Z) would like to express his heartful 
appreciation for the unreserved understandings and supports
from his family.  This work is supported in part by  the NSF of China and 
the National PanDeng (Climb Up) Plan in China. Two of us (Y. Y. and Z. Y. Z)
are also supported by Grant LWTZ-1298 of Chinese Academy of Science.



\begin{references}


\bibitem{Halp} B. I. Halperin, Phys. Rev. B {\bf 25}, 2185 (1982).
\bibitem{Lau} R. B. Laughlin, Phys. Rev. Lett {\bf 50}, 1395 (1983).
\bibitem{Rez} E. H. Rezayi and F. D. M. Haldane, Phys. Rev. B {\bf 50},
6924 (1994).
\bibitem{ECS} F. D. M. Haldane, Bull. Am. Phys. Soc. {\bf 37}, 164(1992);
Phys. Rev. Lett. {\bf 67}, 937 (1991); S. Mitra and A. H. MacDonald, Phys.
Rev. B {\bf 48}, 2005 (1993); P. J. Forrester and B. Jancovici, J. Phys. 
(Pairs) {\bf 45}, L583 (1994); N. Kawakami, Phys. Rev. Lett. 
{\bf 71}, 275(1993); S. Iso and S. J. Rey, cond-mat/9512078; A. D. Veigy 
and S. Ouvry. Phys. Rev. Lett. {\bf 72}, 121(1994).
\bibitem{CFP} L. Brey, Phys. Rev. B {\bf 50}, 11861 (1994);D. B. Chkolvskii,
Phys. Rev. B {\bf 51}, 9895 (1995).
\bibitem{Wen}  X. G. Wen, Phys. Rev. Lett. {\bf 64}, 2206 (1990); Phys.
Rev. B {\bf 41}, 12838 (1990); Int. J. Mod.Phys B, {\bf 6}, 1711 (1992).
\bibitem{DO} Dror Orgad, Phys. Rev. B {\bf 53}, 7964 (1996).
\bibitem{CS}F. Calogero, J. Math.  Phys. {\bf 10} 2197 (1967).
B. Sutherland, J. Math. Phys. {\bf 12}, 246, 251 (1971).  
\bibitem{Yu} Y. Yu and Z. Y. Zhu, preprint, cond-mat/9704124.
\bibitem{Jain} J. K. Jain, Phys. Rev. B {\bf 41}, 7653 (1990) and references 
therein.
\bibitem{LMW} J. M. Leinaas and J. Myrheim, Nuove Cimento {\bf 37} B, 1(1977);
F. Wilczek, Phys. Rev. Lett. {\bf 48}, 1144 (1982).
\bibitem{KHZ} S. C. Zhang, H. Hanson and S. Kivelson, Phys. Rev. Lett.
{\bf 62}, 82 (1989); {\bf 62} 980 (E).
\bibitem{LF} Lopez and E. Fradkin, Phys. Rev. B {\bf 47}, 7080 (1993).
\bibitem{KZ} V. Kalmeyer and S. C. Zhang, Phys. Rev. B {\bf 46}, 9889 (1992).
\bibitem{HLR} B. I. Halperin, P. A. Lee and N. Read, Phys. Rev B {\bf 47}, 
7312 (1993).
\bibitem{RS}R. Rajaraman and S. L. Sondhi, Int. J. Mod. Phys., {\bf B1 10},
793 (1996).
\bibitem{WY} Y. S. Wu and Y. Yu, preprint, cond-mat/9608061.
\bibitem{psudo} F.D. M Haldane, Phy. Rev. Lett. {\bf 51},605 (1983).
\bibitem{KY} N. Kawakami and S.-K. Yang, PRL 67, 2493 (1991).
\bibitem{Wu} Y. S. Wu and Y. Yu, Phys. Rev. Lett. {\bf 75}, 890 (1995).
\bibitem{Wen1} X. G. Wen, Phys. Rev. B {\bf 44}, 5708 (1991).
\bibitem{Ch} D. B. Chklovskii, B. I. Halperin, preprint, cond-mat/9611185.
\bibitem{Hald1} F. D. M. Haldane, J. Phys. C {\bf 14}, 2585 (1981).
\bibitem{WZ} X. G. Wen and A. Zee, Phys. Rev. B {\bf 46 }, 2290 (1992).
\bibitem{EX} See , e. g. Z.N.C. Ha, Nucl. Phys. {\bf435}, 604 (1995).
\bibitem{YY} C. N. Yang and C. P. Yang, J. Math. Phys. {\bf 10}, 1115 (1969).
\bibitem{Frad} E. Fradkin, {\it Field Theories of Condensed Matter Systems},
Addison-Wesley Publishing Company, Redwood City, 1991. 
\bibitem{Hald} F. D. M. Haldane, J. Phys. {\bf C 14}, 2585 (1981).
\bibitem{foot} In an isolated chiral theory coupled to the external fields the 
charge is not conserved because it is known that the theory is anomalous and
not physical \cite{EX}.
\bibitem{LL} See, e. g., L. D. Landau and E. M. Lifshitz, {\it Quantum 
Mechanics (Non-relativistic) }, third edition, Pergamon, Press, Oxford 1987. 


\end{references}
\end{document}